\documentclass[12pt]{article}
\usepackage{axodraw,bbold}

\catcode`@=12
\begin{document}
\thispagestyle{empty}
\begin{flushright} 
UCRHEP-T440\\ 
October 2007\
\end{flushright}
\vspace{0.5in}
\begin{center}
{\LARGE	\bf Multiplicative Conservation of Baryon\\ Number 
and Baryogenesis\\}
\vspace{1.5in}
{\bf Ernest Ma\\}
\vspace{0.2in}
{\sl Department of Physics and Astronomy, University of California,\\} 
\vspace{0.1in}
{\sl Riverside, California 92521, USA\\}
\vspace{1.5in}
\end{center}

\begin{abstract}\
In the canonical seesaw mechanism of neutrino mass, lepton number is only 
multiplicatively conserved, which enables the important phenomenon of 
leptogenesis to occur, as an attractive explanation of the present baryon 
asymmetry of the Universe.  A parallel possibility, hitherto unrecognized, 
also holds for baryon number and baryogenesis.  This new idea is shown to be 
naturally realized in the context of a known supersymmetric string-inspired 
extension of the Standard Model, based on $E_6$ particle content, and having 
an extra $U(1)_N$ gauge symmetry.  Within this framework, two-loop radiative 
neutrino masses are also possible, together with a new form of very long-lived 
matter.
\end{abstract}

\newpage
\baselineskip 24pt

The Universe has an imbalance of matter over antimatter, called the baryon 
asymmetry \cite{c06}.  An elegant way of understanding this is the phenomenon 
of leptogenesis \cite{fy86,bpy05}, by which a lepton asymmetry is established 
from the decays of heavy Majorana singlet fermions $N_i$ and gets converted 
\cite{krs85} into a baryon asymmetry through sphalerons \cite{km84} during 
the electroweak phase transition in the early Universe.  In this scenario, 
lepton number is only multiplicatively conserved and neutrinos acquire 
small Majorana masses through the famous canonical seesaw mechanism 
\cite{seesaw}.

In this paper, a parallel and equally elegant possibility, i.e. 
multiplicatively conserved baryon number and baryogenesis, is proposed 
and shown to be naturally realized in the framework of a known 
supersymmetric string-inspired extension \cite{m96} of the Standard Model 
(SM), as detailed below.

In leptogenesis, the only interactions of $N_i$ are with the lepton doublets 
$(\nu_i,l_i)$ and the Higgs doublet $(\phi^+,\phi^0)$.  As the Universe 
expands and cools, the out-of-equilibrium decays of $N_1$ (i.e. the lightest 
$N_i$) into $l^- \phi^+$ and $l^+ \phi^-$ establish a lepton ($L$) asymmetry. 
This engenders a baryon ($B$) asymmetry through sphaleron interactions which 
change $B+L$ but not $B-L$.

Consider now the following extension of the SM.  Let 
$\tilde h$, $\tilde h^c$ be heavy singlet scalar quarks of charge $Q = 
\mp 1/3$ and baryon number $B = \mp 2/3$; and let $N^c_{1,2}$ be heavy 
singlet neutral fermions of $B=1$.  As a result, the new interactions 
\begin{equation}
Q Q \tilde h, ~~~ u^c d^c \tilde h^c, ~~~ d^c N^c \tilde h, 
\end{equation}
where $Q = (u,d)$, are allowed. Suppose also that $N^c_{1,2}$ are Majorana 
so that baryon number is only multiplicatively conserved.  Then the decays
\begin{eqnarray}
N^c_1 &\to& \bar{d}^c \tilde{h}^* \to \bar{d}^c d u ~~ (B=+1), \\ 
N^c_1 = \bar{N}^c_1 &\to& d^c \tilde h \to d^c \bar d \bar u ~~ (B=-1),
\end{eqnarray}
generate a baryon asymmetry under the same conditions as in leptogenesis. 
Again $N^c_2$ is needed to obtain the required CP violation from the 
interference of the tree and one-loop diagrams as shown in Fig.~1.

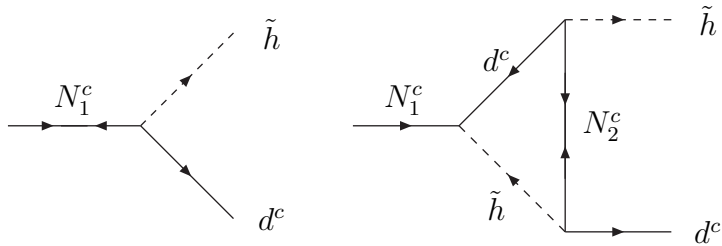
\begin{figure}[htb]
\begin{center}
\begin{picture}(360,100)(0,10)
\ArrowLine(50,50)(80,50)
\ArrowLine(100,50)(70,50)
\ArrowLine(100,50)(135,15)
\DashArrowLine(100,50)(135,85)3

\ArrowLine(180,50)(220,50)
\ArrowLine(260,90)(220,50)
\DashArrowLine(260,10)(220,50)3
\ArrowLine(260,90)(260,30)
\ArrowLine(260,10)(260,70)
\ArrowLine(260,10)(300,10)
\DashArrowLine(260,90)(300,90)3

\Text(75,60)[]{$N_1^c$}
\Text(150,85)[]{$\tilde{h}$}
\Text(150,15)[]{$d^c$}
\Text(200,60)[]{$N_1^c$}
\Text(235,75)[]{$d^c$}
\Text(235,20)[]{$\tilde{h}$}
\Text(275,50)[]{$N_2^c$}
\Text(315,90)[]{$\tilde{h}$}
\Text(315,10)[]{$d^c$}

\end{picture}
\end{center}
\caption{Tree and one-loop diagrams for $N^c_1 \to d^c \tilde{h}$.}
\end{figure}

\noindent As an illustration, let $m_{N^c} \sim 10^6$ GeV and $d^c N^c_2 
\tilde{h}$ couplings $\sim 10^{-2}$, then a decay asymmetry of order $10^{-6}$ 
may be established, enabling a baryon asymmetry $\eta_B \sim 10^{-10}$ to be 
obtained.  The out-of-equilibrium condition requires however that the 
$d^c N^c_1 \tilde{h}$ couplings be less than about $10^{-5}$. Sphaleron 
interactions will modify this pure $B$ asymmetry into a $B-L$ asymmetry 
in exact analogy to what happens to a pure $L$ asymmetry in the case of 
leptogenesis.

This new idea of multiplicatively conserved baryon number and baryogenesis 
turns out to be naturally realized in the context of a supersymmetric 
string-inspired extension of the SM proposed some time ago \cite{m96}, 
with many interesting features in its own right \cite{many,ms07}.   
It is based on $E_6$ with matter content given by three \underline{27} 
representations and with gauge interactions of the SM  plus those of 
$U(1)_N$, which is a linear combination of $U(1)_\psi$ and $U(1)_\chi$ 
in the decomposition:
\begin{eqnarray}
E_6 &\to& SO(10) \times U(1)_\psi, \\
SO(10) &\to& SU(5) \times U(1)_\chi.
\end{eqnarray}
In terms of the maximal subgroup $SU(3)_C \times SU(3)_L \times SU(3)_R$ of 
$E_6$, the $U(1)_N$ charge is given by
\begin{equation}
Q_N = 6 Y_L + T_{3R} - 9 Y_R,
\end{equation}
where $T_{3L,3R}$ and $Y_{L,R}$ are the usual quantum numbers of the 
$SU(2) \times U(1)$ decompositions of $SU(3)_{L,R}$. The particle content 
of a \underline{27} multiplet of $E_6$ is shown in Table 1.
 
\begin{table}[htb]
\caption{Particle content of \underline{27} of $E_6$ under $SU(3)_C \times 
SU(2)_L \times U(1)_Y$ and $U(1)_N$.}
\begin{center}
\begin{tabular}{|c|c|c|}
\hline 
Superfield & $SU(3)_C \times SU(2)_L \times U(1)_Y$ & $U(1)_N$ \\ 
\hline
$Q = (u,d)$ & (3,2,1/6) & 1 \\
$u^c$ & $(3^*,1,-2/3)$ & 1 \\ 
$e^c$ & (1,1,1) & 1 \\
\hline
$d^c$ & $(3^*,1,1/3)$ & 2 \\ 
$L = (\nu,e)$ & $(1,2,-1/2)$ & 2 \\ 
\hline
$h$ & $(3,1,-1/3)$ & $-2$ \\ 
$\bar{E} = (E^c,N^c_E)$ & $(1,2,1/2)$ & $-2$ \\ 
\hline
$h^c$ & $(3^*,1,1/3)$ & $-3$ \\ 
$E = (\nu_E,E)$ & $(1,2,-1/2)$ & $-3$ \\ 
\hline
$S$ & $(1,1,0)$ & 5 \\
\hline
$N^c$ & (1,1,0) & 0 \\ 
\hline
\end{tabular}
\end{center}
\end{table}

\noindent There are eleven possible generic trilinear terms invariant under 
$SU(3)_C \times SU(2)_L \times U(1)_Y \times U(1)_N$.  Five are necessary 
for fermion masses, namely
\begin{equation}
Qu^c\bar{E}, ~~ Qd^cE, ~~ Le^cE, ~~ SE\bar{E}, ~~ Shh^c,
\end{equation}
for $m_u$, $m_d$, $m_e$, $m_E$, $m_h$ respectively. The other six are
\begin{equation}
LN^c\bar{E}, ~~ QLh^c, ~~ u^ce^ch, ~~ d^cN^ch, ~~ QQh, ~~ u^cd^ch^c,
\end{equation}
some of which must be absent to prevent rapid proton decay.  Hence all such 
models require an additional discrete symmetry, the simplest of which is 
of course a single $Z_2$, resulting in eight generic possibilities, as 
shown already many years ago \cite{m88}.  I consider here instead 
an exactly conserved $(-)^L \times (-)^{3B}$ symmetry as shown in Table 2.

\begin{table}[htb]
\caption{Particle content of \underline{27} of $E_6$ under $(-)^L$   
and $(-)^{3B}$.}
\begin{center}
\begin{tabular}{|c|c|c|}
\hline 
Superfield & $(-)^L$ & $(-)^{3B}$ \\ 
\hline
$Q,u^c,d^c$ & + & $-$ \\
$L,e^c$ & $-$ & + \\ 
$h,h^c$ & + & + \\
$E,\bar{E},S$ & $+$ & + \\ 
$N^c$ & $+$ & $-$ \\ 
\hline
\end{tabular}
\end{center}
\end{table}

As a result, all terms of Eq.~(7) are allowed as well as the following from 
Eq.~(8):
\begin{equation}
QQh, ~~~ u^c d^c h^c, ~~~ d^c N^c h,
\end{equation}
exactly as desired, i.e. those of Eq.~(1).  In addition, $N^c$ are allowed 
large Majorana masses, and the undesirable bilinear terms $L \bar{E}$ and 
$d^c h$ (allowed by $U(1)_N$ alone) are forbidden by $(-)^L$ and $(-)^{3B}$ 
respectively.  Successful baryogenesis is therefore possible.  Note that 
below the $N^c$ mass scale, the theory is both $B$ and $L$ conserving with 
$h,h^c$ having $B = \mp 2/3$.  This is important so that the only violation 
($B+L$) comes from the sphalerons.

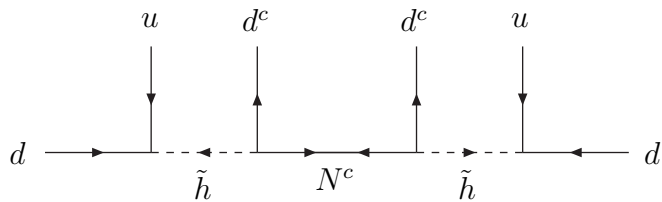
\begin{figure}[htb]
\begin{center}
\begin{picture}(360,80)(0,10)
\ArrowLine(70,20)(110,20)
\ArrowLine(110,60)(110,20)
\ArrowLine(150,20)(150,60)
\DashArrowLine(150,20)(110,20)3

\ArrowLine(150,20)(190,20)
\ArrowLine(210,20)(170,20)
\DashArrowLine(210,20)(250,20)3
\ArrowLine(210,20)(210,60)
\ArrowLine(250,60)(250,20)
\ArrowLine(290,20)(250,20)

\Text(60,20)[]{$d$}
\Text(110,70)[]{$u$}
\Text(150,70)[]{$d^c$}
\Text(210,70)[]{$d^c$}
\Text(250,70)[]{$u$}
\Text(300,20)[]{$d$}
\Text(130,8)[]{$\tilde{h}$}
\Text(230,8)[]{$\tilde{h}$}
\Text(180,10)[]{$N^c$}

\end{picture}
\end{center}
\caption{Diagram for deuteron decay and neutron-antineutron oscillation.}
\end{figure}

\noindent Since $(-)^{3B}$ is conserved, the proton is stable, but the 
deuteron is not, and neutron-antineutron oscillations are allowed.  The 
latter two processes are based on the same mechanism responsible for 
baryogenesis, i.e. Eqs.~(2) and (3), as shown in Fig.~2.
As an illustration, let $m_{N^c} \sim 10^6$ GeV, $m_{\tilde{h}} \sim 10^3$ 
GeV, $d^c N^c \tilde{h}$ couplings $\sim 10^{-2}$, and $u d \tilde{h}$ 
couplings $\sim 10^{-5}$, then $(10^{-2})^2 (10^{-5})^2/(10^6)(10^3)^4 = 
(10^{-16})^2$, implying an equivalent scale of $10^{16}$ GeV in the usual 
consideration of baryon-number nonconserving processes at the GeV scale, 
in agreement with all present experimental bounds.  As for $R$ parity, it 
remains the same, i.e. $R = (-)^{3B+L+2j}$, in this model as in the Minimal 
Supersymmetric Standard Model (MSSM).  The lightest neutralino is thus 
again a good candidate for the dark matter of the Universe.

Since $N^c$ is odd under $(-)^{3B}$ but even under $(-)^L$, there is no 
$L N^c \bar{E}$ coupling and $N^c$ does not play the role of a singlet 
right-handed neutrino as is normally assumed. This means that neutrino 
masses remain zero, contrary to present data on neutrino oscillations. 
To remedy this shortcoming, a very interesting variation of the above 
scenario is described below.

\begin{table}[htb]
\caption{Division of $E, \bar{E}, S$ superfields under $(-)^L$   
and $(-)^{3B}$.}
\begin{center}
\begin{tabular}{|c|c|c|}
\hline 
Superfield & $(-)^L$ & $(-)^{3B}$ \\ 
\hline
$E_1,\bar{E}_1,S_1$ & $+$ & + \\ 
$E_{2,3},\bar{E}_{2,3},S_{2,3}$ & $-$ & $-$ \\ 
\hline
\end{tabular}
\end{center}
\end{table}

There are three sets of $E, \bar{E}, S$ superfields, but only one is 
required to break $SU(2)_L \times U(1)_Y \times U(1)_N$ and to endow all 
fermions (except the neutrinos) with masses. Let them thus be divided as 
in Table 3. In that case, the generic $S E \bar{E}$ couplings are restricted 
to $S_1 E_{1,2,3} \bar{E}_{1,2,3}, ~ S_{2,3} E_1 \bar{E}_{2,3}, ~ S_{2,3} 
E_{2,3} \bar{E}_1$, and the important new couplings
\begin{equation}
L_i N^c_j \bar{E}_{2,3}
\end{equation}
are allowed.  The decays of $N^c_1$ into $l^- \tilde{E}^+$ and $l^+ 
\tilde{E}^-$ now also generate a lepton asymmetry. Thus remarkably, both 
$B$ and $L$ asymmetries are established in the decays of $N^c_1$, and at 
energies below its mass, the theory conserves both $B$ and $L$. 
Sphaleron interactions then convert both asymmetries into a $B-L$ 
asymmetry, the baryon component of which is observed today.

\begin{figure}[htb]
\begin{center}
\begin{picture}(360,120)(0,0)
\ArrowLine(100,10)(140,10)
\ArrowLine(190,10)(140,10)
\ArrowLine(170,10)(220,10)
\ArrowLine(260,10)(220,10)
\DashArrowLine(173,43)(140,10)3
\DashArrowLine(187,43)(220,10)3
\DashArrowLine(150,80)(173,57)3
\DashArrowLine(210,80)(187,57)3
\CArc(180,50)(10,0,360)

\Text(120,0)[]{$\nu_i$}
\Text(240,0)[]{$\nu_j$}
\Text(180,-2)[]{$N^c_k$}
\Text(145,40)[]{$(\tilde{N}^c_E)_{2,3}$}
\Text(215,40)[]{$(\tilde{N}^c_E)_{2,3}$}
\Text(130,85)[]{$(\tilde{N}^c_E)_1$}
\Text(230,85)[]{$(\tilde{N}^c_E)_1$}

\end{picture}
\end{center}
\caption{Two-loop generation of neutrino mass.}
\end{figure}
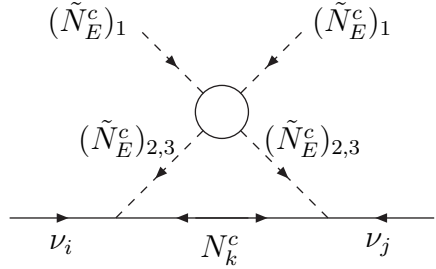

Using the interactions of Eq.~(10) and the one-loop effective 
$[(\tilde{N}^c_E)_1^\dagger (\tilde{N}^c_E)_{2,3}]^2$ couplings from 
supersymmetry breaking, seesaw neutrino masses are now generated in two 
loops as shown in Fig.~3. This has the same structure discussed 
in Ref.~\cite{ms07}.  However, no extra $Z_2$ symmetry beyond $(-)^L$ and 
$(-)^{3B}$ is assumed here. For $N^c$ of order $10^6$ GeV, realistic 
neutrino masses of order 0.1 eV may be obtained.  This is well below the 
bound of $10^9$ GeV on the reheating temperature of the Universe for 
avoiding the overproduction of gravitinos \cite{gravitino}.

Since $(-)^{3B+L}$ is still the same, i.e. even, for all $E, \bar{E}, S$ 
superfields, $R$ parity is unchanged in this scenario.  However, the 
lightest particle contained in $E_{2,3}$, $\bar{E}_{2,3}$, and $S_{2,3}$ 
must decay through $N^c$, so its lifetime is very long, say of order 
$10^6$ seconds. An example of such a decay is shown in Fig.~4.
Depending on their masses, there may even be two such long-lived particles, 
one with $R$ parity even and the other odd.

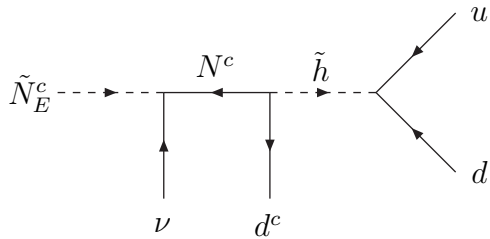
\begin{figure}[htb]
\begin{center}
\begin{picture}(360,80)(0,0)
\DashArrowLine(100,50)(140,50)3
\ArrowLine(140,10)(140,50)
\ArrowLine(180,50)(140,50)
\DashArrowLine(180,50)(220,50)3
\ArrowLine(180,50)(180,10)
\ArrowLine(250,80)(220,50)
\ArrowLine(250,20)(220,50)

\Text(90,50)[]{$\tilde{N}^c_E$}
\Text(140,0)[]{$\nu$}
\Text(180,0)[]{$d^c$}
\Text(160,60)[]{$N^c$}
\Text(200,60)[]{$\tilde{h}$}
\Text(260,80)[]{$u$}
\Text(260,20)[]{$d$}

\end{picture}
\end{center}
\caption{Diagram for the long-lived decay of $\tilde{N}^c_E$.}
\end{figure}

\newpage

To summarize, the new idea of multiplicative conservation of baryon number 
and baryogenesis has been proposed.

(1) It requires at the minimum the existence of a scalar singlet 
quark $\tilde{h}$ of charge $Q=-1/3$ and baryon number $B=-2/3$.  If 
kinematically allowed, $\tilde{h} \tilde{h}^*$ will be produced at the 
Large Hadron Collider (LHC), and identified through their subsequent decays 
into 4 quark jets with large transverse momenta and large pairwise 
invariant masses.

(2) The lightest of at least 2 heavy neutral singlet Majorana 
fermions $N^c_i$ may then decay into $\tilde{h}^* \bar{d}^c~(B=+1)$ and 
$\tilde{h} d^c~(B=-1)$ and establish a baryon asymmetry of the Universe. 
The mass of $N^c_1$ (the lightest $N^c_i$) may be of order $10^6$ GeV.

(3) The proton is stable, but the deuteron is not, and 
neutron-antineutron oscillations are allowed.

(4) This scenario is naturally realized in a known 
supersymmetric string-inspired extension of the SM, i.e. $SU(3)_C \times 
SU(2)_L \times U(1)_Y \times U(1)_N$ with particle content given by three 
\underline{27} representations of $E_6$.

(5) If kinematically allowed, the $U(1)_N$ gauge boson $Z_N$ will 
be discovered with ease at the LHC because it has both quark and lepton 
couplings (see Table 1).

(6) Two-loop radiative seesaw neutrino masses are also possible in 
an interesting variation of the model, where both $B$ and $L$ asymmetries are 
established by the decays of $N^c_1$, to be converted into a $B-L$ asymmetry 
by sphaleron interactions.

(7) The lightest particle odd under $R=(-)^{3B+L+2j}$ is a 
candidate for the dark matter of the Universe as in the MSSM.  However, 
there is now at least one particle which is very long-lived, and will also 
appear as missing energy at the LHC.

This work was supported in part by the U.~S.~Department of Energy under
Grant No. DE-FG03-94ER40837. 

\newpage

\baselineskip 18pt

\bibliographystyle{unsrt}

\end{document}